\documentclass[twocolumn,tighten]{aastex62}

\shorttitle{Model-independent test of isotropy with Pantheon SN}
\shortauthors{Andrade et al.}

\begin{document}

\title{A MODEL-INDEPENDENT TEST OF COSMIC ISOTROPY WITH LOW-Z PANTHEON SUPERNOVAE}

\correspondingauthor{Uendert Andrade}
\email{uendertandrade@on.br}

\author{Uendert Andrade}
\affiliation{Observat\'orio Nacional 20921-400, Rio de Janeiro, RJ, Brasil}

\author{Carlos A. P. Bengaly}
\affiliation{Department of Physics \& Astronomy, University of the Western Cape, 7535, Cape Town, South Africa}

\author{Beethoven Santos}
\affiliation{Observat\'orio Nacional 20921-400, Rio de Janeiro, RJ, Brasil}

\author{Jailson S. Alcaniz}
\affiliation{Observat\'orio Nacional 20921-400, Rio de Janeiro, RJ, Brasil}
\affiliation{Departamento de F\'{\i}sica, Universidade Federal do Rio Grande do Norte, 59072-970, Natal, RN, Brasil}

\begin{abstract}

The assumption of homogeneity and isotropy on large scales is one of the main hypothesis of the standard cosmological model. In this paper, we revisit a test of cosmological isotropy using type Ia supernova (SN Ia) distances provided by the latest SN Ia compilation available, namely, the Pantheon compilation. We perform a model-independent analysis by selecting low-redshift subsamples lying in two redshift intervals, i.e., $z \leq 0.10$ and $z \leq 0.20$. By mapping the directional asymmetry of cosmological parameters across the sky, we show that the current SN Ia data favours the hypothesis of cosmic isotropy, as the anisotropy found in the maps can be mostly ascribed to the non-uniform sky coverage of the data rather than an actual cosmological signal. These results confirm that there is null evidence against the cosmological principle in the low-redshift universe. 
\end{abstract}

\keywords{Cosmology: Theory -- Cosmological Principle -- Distance Scale}

\section{Introduction} \label{sec:intro}

The assumption of cosmological homogeneity and isotropy, i.e., the Cosmological Principle (CP), constitutes a fundamental pillar of modern Cosmology (see, e.g., \citealt{1995PhRvD..52.1821G, 2010CQGra..27l4008C}) and implies that cosmic distances and ages can be directly derived from the Friedmann-Lema\^itre-Robertson-Walker (FLRW) metric. Given that any violation of the CP would have major consequences to our description of the Universe, it is of crucial importance to determine whether it actually holds true in light of observational data. If not, the current concordance model of Cosmology ($\Lambda$CDM), for instance, would need to be profoundly revised and reformulated.  

One of the most direct ways to probe isotropy is through the angular dependence of the Type Ia Supernovae (SN Ia) Hubble diagram\footnote{For recent measurements of the cosmic homogeneity scale we refer the reader to~\cite{2017JCAP...06..019N, 2018MNRAS.475L..20G}.}. Using this information, we can assess how the best-fit cosmological parameters vary across the sky. Any hint of large-scale departure from isotropy would be revealed if these maps disagree with synthetic data-sets based on the concordance model, or idealised sky distribution of data points. Thus far, most of analyses performed with SNe Ia data showed good agreement with the isotropy assumption~\citep{2001MNRAS.323..859K, 2007A&A...474..717S, 2010JCAP...12..012A, 2011MNRAS.414..264C, 2012JCAP...02..004C, 2013A&A...553A..56K, 2013PhRvD..87l3522C, 2015ApJ...801...76A, 2015ApJ...808...39B, 2015ApJ...810...47J, 2016MNRAS.456.1881L, 2016JCAP...04..036B, 2018PhRvD..97h3518A, 2018PhRvD..97l3515D, 2018arXiv180602773D, 2018arXiv180405191S, 2018MNRAS.478.5153S}. Nevertheless, potential indications of isotropy departure were found at CMB latest data~\citep{2016A&A...594A..16P, 2016CQGra..33r4001S}, as well as in the dipole anisotropy of radio source number counts~\citep{2011ApJ...742L..23S, 2018JCAP...04..031B}. Hence, further investigations on this subject are still needed. 

In this work, we revisit the isotropy test with cosmic distances, as performed in~\citep{2015ApJ...808...39B, 2016JCAP...04..036B}, with the latest SN Ia compilation, the so-called Pantheon compilation~\citep{2018ApJ...859..101S}. Some recent analyses confirmed previous results, i.e., null evidence for isotropy violation from these data~\citep{2018MNRAS.478.5153S, 2018arXiv180602773D}. However, these works assumed {\it a priori} the concordance model to describe the SN Ia distances, thus testing consistency with it. We perform our analyses in a low-$z$ range instead, which allows us to do it in a {\it model-independent} way using a cosmographic approach~\citep{2004CQGra..21.2603V}. We found that the celestial anisotropy of cosmographic parameters is mostly due to the non-uniform distribution of SN Ia data points, and that it displays excellent concordance with Monte Carlo simulations based on the concordance model.


\begin{figure*}
\includegraphics[width=0.30\textwidth, height=0.38\textheight, angle=90]{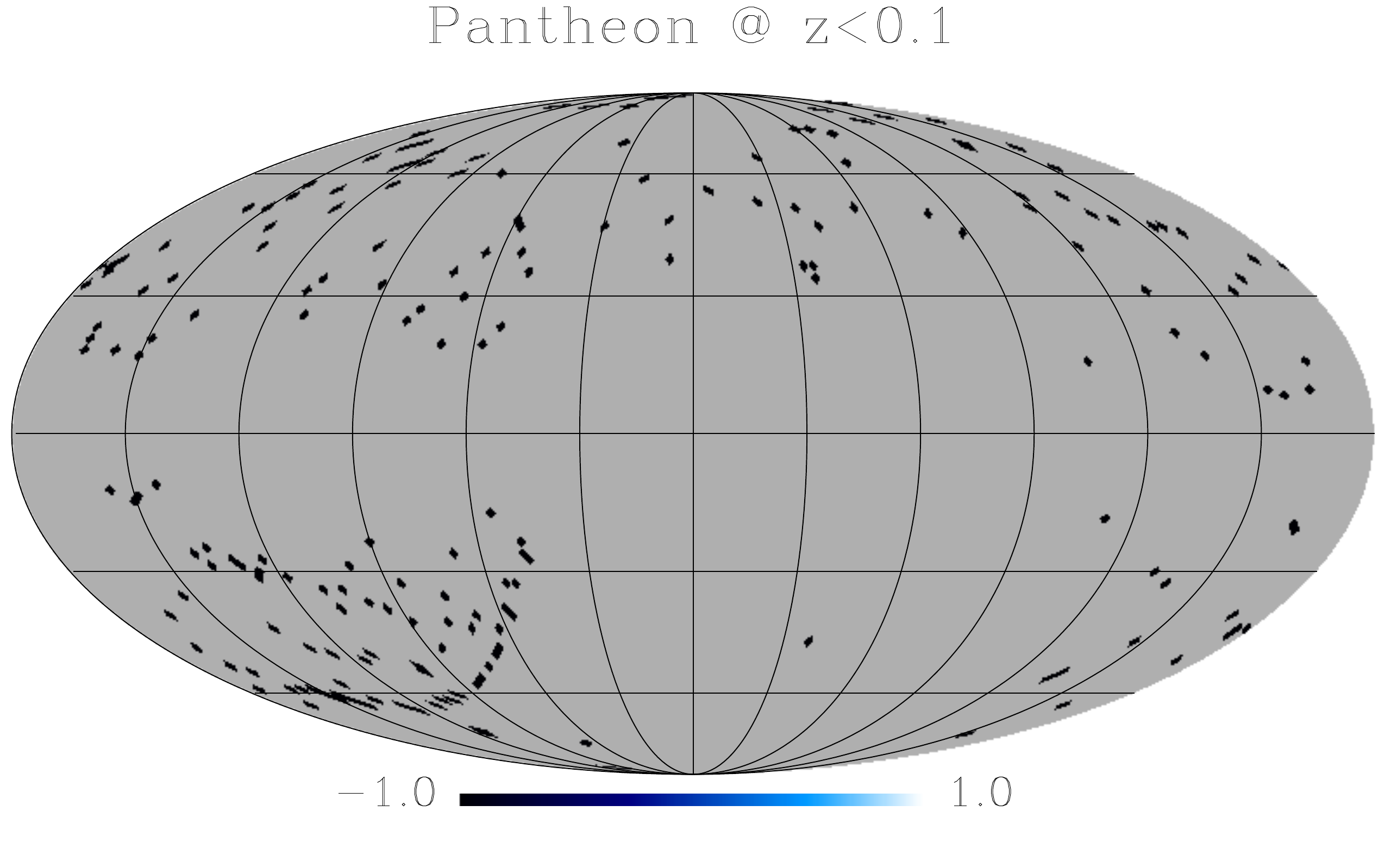}
\includegraphics[width=0.30\textwidth, height=0.38\textheight, angle=90]{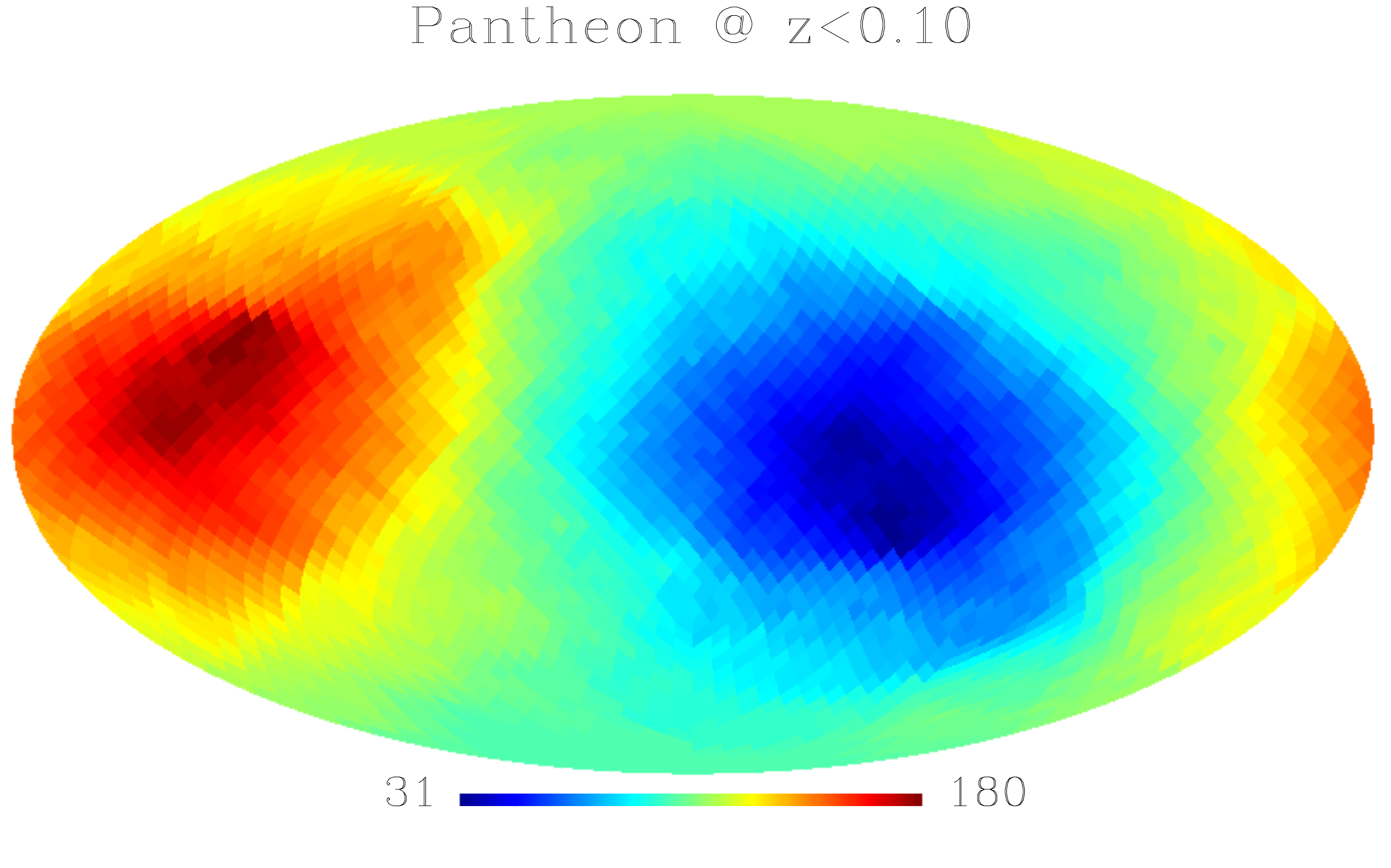}
\caption{Left panel: The sky distribution of Pantheon SN Ia at $z \leq 0.10$. Right panel: The number of objects encompassed in each hemisphere selected for our analyses (see text for description).}
\label{fig:SN_maps_z0p1}
\end{figure*}

\begin{figure*}
\includegraphics[width=0.30\textwidth, height=0.38\textheight, angle=90]{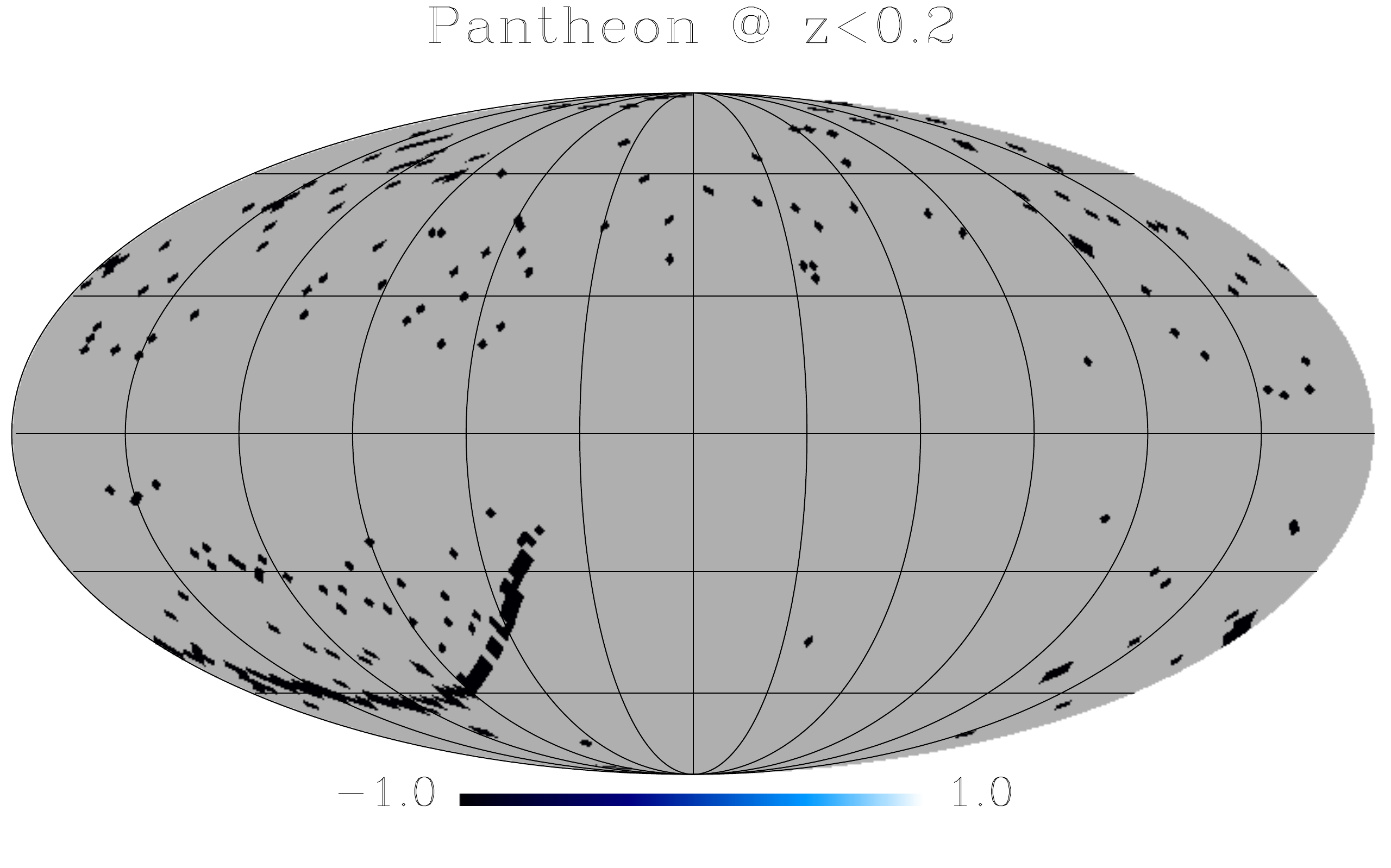}
\includegraphics[width=0.30\textwidth, height=0.38\textheight, angle=90]{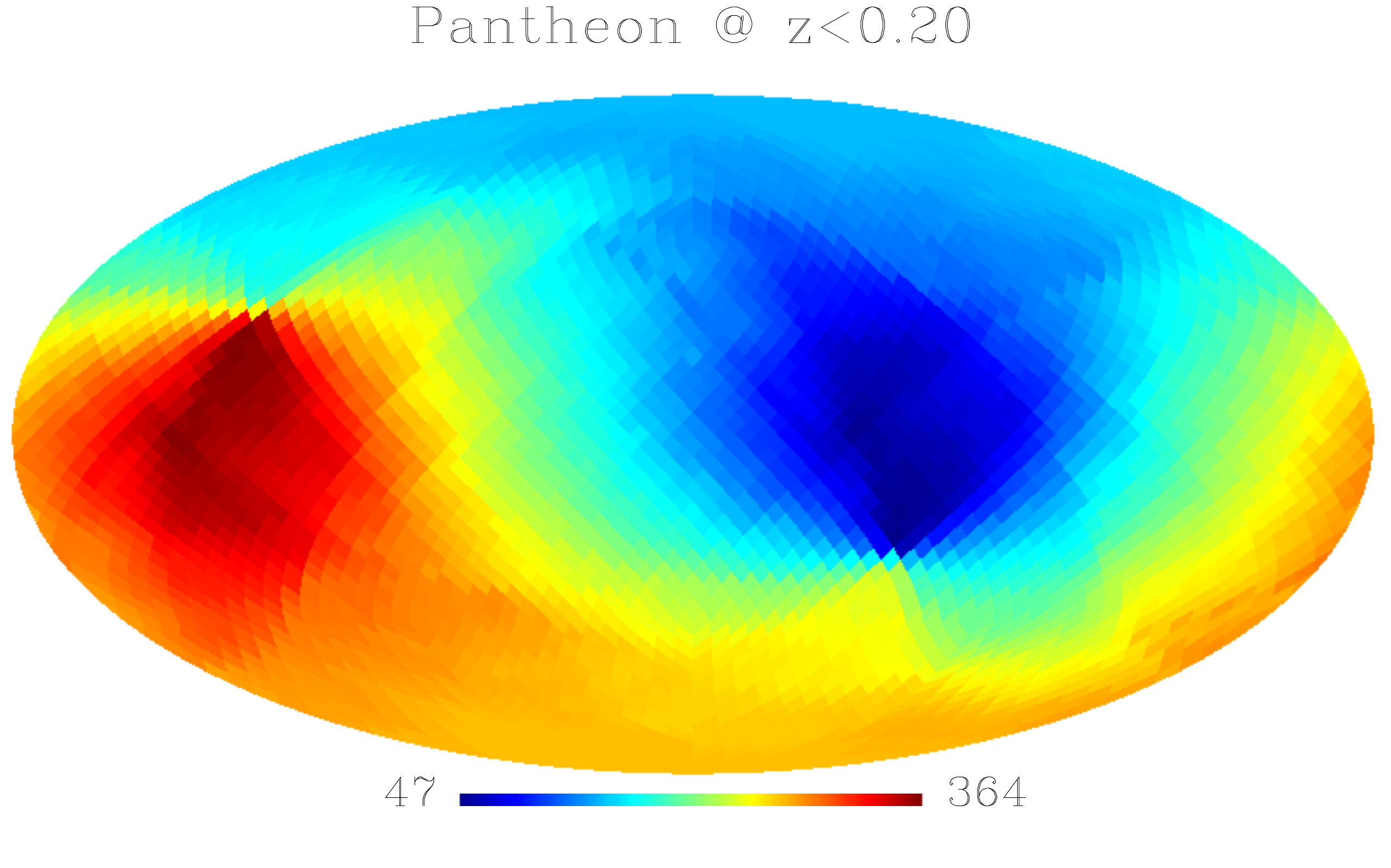}
\caption{Same as Figure~\ref{fig:SN_maps_z0p1}, but for the SN Ia lying in the $z \leq 0.20$ interval instead.}
\label{fig:SN_maps_z0p2}
\end{figure*}

\section{Observational data and method} \label{sec:data_method}

In our analysis, we adopt the Pantheon SN Ia compilation, which consists of the largest and most complete SN data-set at the present moment, i.e., 1049 objects lying in the interval $0.01 < z < 2.30$ compiled from the PanSTARRS1 Medium Deep Survey, SDSS, SNLS, in addition to many low-$z$ and HST data points. As we will focus on a model-independent analysis, we will only select objects at $z \leq 0.10$ and $z \leq 0.20$, thus reducing our sample to 211 and 411 data points, respectively\footnote{For the sake of comparison, older SN compilations, such as Union2.1 and JLA, encompass 211 and 317 data points at $z \leq 0.20$, respectively.}. The sky distribution of the selected SN Ia at $z \leq 0.10$ can be visualized on the left panel of Fig.~\ref{fig:SN_maps_z0p1}, whereas the left panel of Fig.~\ref{fig:SN_maps_z0p2} shows the SNe Ia at $z \leq 0.20$. On the other hand, the right panel of both figures display the asymmetry in the number of SN Ia in hemispheres whose symmetry axes coincide pixel centers defined according to {\sc HEALPix}~\citep{2005ApJ...622..759G} $N_{\rm side}=16$ grid resolution. Then, we can note how non-uniform are the SN celestial distributions at both redshift ranges, as the number of objects varies from 31 to 180 across the sky at $z \leq 0.10$, and from 47 to 384 at $z \leq 0.20$.    

Assuming that the FLRW metric holds true, one can expand the scale factor around the present time, and then measure distances regardless of the Universe dynamics. This is the well-known cosmographic approach (see, e.g.~\cite{1972gcpa.book.....W}). The luminosity distance reads
\begin{eqnarray}\label{eq:DL_z}
D_{\rm L}(z) = 3000h^{-1}_0 \left[z + (1 - q_0)(z^2/2)  + O(z^3)\right] \;,
\end{eqnarray}
where $z$ is the redshift observed in the comoving rest frame with respect to the expansion of the Universe, $h_0$ and $q_0$ are the dimensionless Hubble constant and decelerating parameter at present time, respectively\footnote{The present value of the Hubble parameter is written as  $H_0 = 100h_0$ $\rm{km.s^{-1}.Mpc^{-1}}$.}, for $D_{\rm L}(z)$ given in Mpc. In order to avoid the issue of divergences up to higher redshifts, we parametrize the redshift variable by $y \equiv z/(1 + z)$~\citep{2007gr.qc.....3122C}. Therefore, we can rewrite the luminosity distance in terms of $y$ such as
\begin{eqnarray}\label{eq:DL_y}
D_{\rm L} (y)  = 3000h^{-1}_0 \left[y + (3 - q_0)(y^2/2) + O(y^3)\right]  \;.
\end{eqnarray}
Therefore, the distance modulus can be written as
\begin{eqnarray}\label{eq:mu}
\mu(y) = 5\log_{10}\left({D_{\rm L}(y)/\rm{Mpc}}\right) + 25 \;.
\end{eqnarray}
As shown in Eq.~\ref{eq:DL_y}, the luminosity distance depends only on $h_0$ and $q_0$ up to the second order in redshift. Therefore, we restrict our analysis up to that order. As shown in~\cite{2015ApJ...808...39B}, this truncation does not bias $\mu$ in the interval $z \lesssim 0.2$.  

We probe the isotropy of the Pantheon SN Ia distances by mapping the directional dependence of $h_0$ and $q_0$ and considering two redshift intervals, i.e., $z \leq 0.10$ and $z \leq 0.20$, similarly to previous analyses in the literature~\citep{2001MNRAS.323..859K, 2007A&A...474..717S, 2010JCAP...12..012A, 2013A&A...553A..56K, 2015ApJ...808...39B, 2015ApJ...810...47J, 2016JCAP...04..036B}. This is done by defining hemispheres whose symmetry axes are given by {\sc HEALPix}~\citep{2005ApJ...622..759G} pixel centers at $N_{\rm side}=16$ grid resolution, and then estimating the $h_0$ and $q_0$ best-fits for all SN Ia enclosed in such hemispheres. To do so, we minimize the following quantity
\begin{equation}\label{eq:chi2}
\chi^2 = \sum_i \left[ ( \mu^{\rm th}(\mathbf{p},y_{\rm i})-\mu^{\rm obs}_{\rm i} )^2/\sigma_{\rm i} \right]^2 \;,
\end{equation}
when fitting each parameter $\mathbf{p}$, being $\mathbf{p} = h_0$ or $q_0$ in our analysis\footnote{When we fit $h_0$, $q_0$ is fixed at $q_0 = -0.574$, which is fully consistent with $\Omega_{\rm m}=0.274$, i.e., Pantheon best fit for flat $\Lambda$CDM. On the other hand, $h_0$ is set to $h_0=0.678$ when $q_0$ is fitted, thus consistent with Planck's best fit for the same model. We checked different values of $h_0$ and $q_0$ as well, but our results were mostly unchanged. A similar procedure was adopted in~\cite{2013A&A...553A..56K}.}. In the above equation, $i$ represents the $i$-th data point belonging to each hemisphere, $\mu^{\rm obs}_{\rm i}$ denotes its distance modulus, $\sigma^2_{\rm i}$ corresponds to its respective uncertainty, and $\mu^{\rm th}(\mathbf{p},y_{\rm i})$ is the theoretically expected distance modulus calculated according to Eq.~(\ref{eq:mu}). We only use the statistical errors for $\mu^{\rm obs}_{\rm i}$ in our analyses, as the full covariance matrix would significantly degrade the constraints at such low-$z$ ranges. Hence, we obtain $3072$ values of $h_0$ and $q_0$ across the entire sky, which will be hereafter referred to as Hubble-maps and $q$-maps, respectively.

The statistical significance of these maps is computed from two sets of 1000 Monte Carlo (MC) realizations according to the following prescriptions: 

\begin{itemize}

\item {\it MC-iso}: The SN original positions in the sky are changed according to an isotropic distribution;

\item {\it MC-lcdm}: The SN original distance moduli are changed according to a value drawn from a normal distribution $\mathcal{N}(\mu_{\rm fid},\sigma)$, i.e., a distribution centered at $\mu_{\rm fid}$, which is fixed at a fiducial Cosmology following $h_0=0.678$ and $q_0=-0.574$, and whose standard deviation is given by the original distance modulus uncertainty, $\sigma$.  

\end{itemize}

We quantify the level of anisotropy in the data and the MCs by defining  
\begin{equation}\label{eq:delta}
\Delta p \equiv p^{\rm max}_0 - p^{\rm min}_0
\end{equation}

\noindent being $p^{\rm max}_0$ and $p^{\rm min}_0$, respectively, the maximum and minimum best fits for $p \rightarrow h_0$ or $q_0$ obtained across the entire celestial sphere. 

Thus, we compare the $\Delta h$ and $\Delta q$ between the real data and these 1000 MCs by computing the fraction of realizations with $\Delta h$ or $\Delta q$ at least as great as the observed one, hence defined as our $p$-value. If we find less than $5\%$ of agreement between them, we can state that its anisotropy can be either ascribed to the non-uniform celestial distribution of SN Ia (for the {\it MC-iso}), or to a departure of the concordance model (for the {\it MC-lcdm}). 

\section{Results} \label{sec:results}

\begin{figure*}[t]
\includegraphics[width=0.30\textwidth, height=0.38\textheight, angle=90]{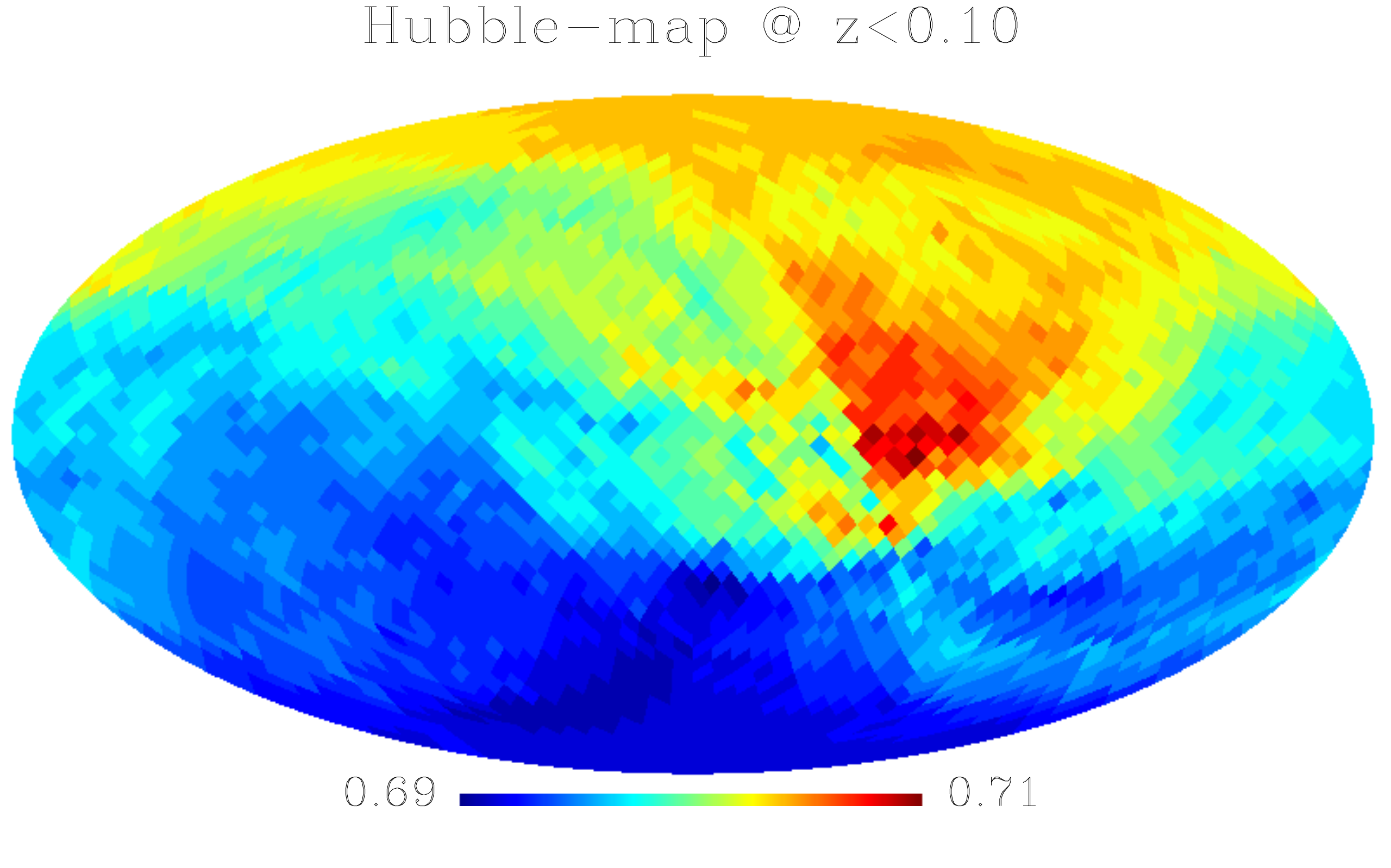}
\includegraphics[width=0.30\textwidth, height=0.38\textheight, angle=90]{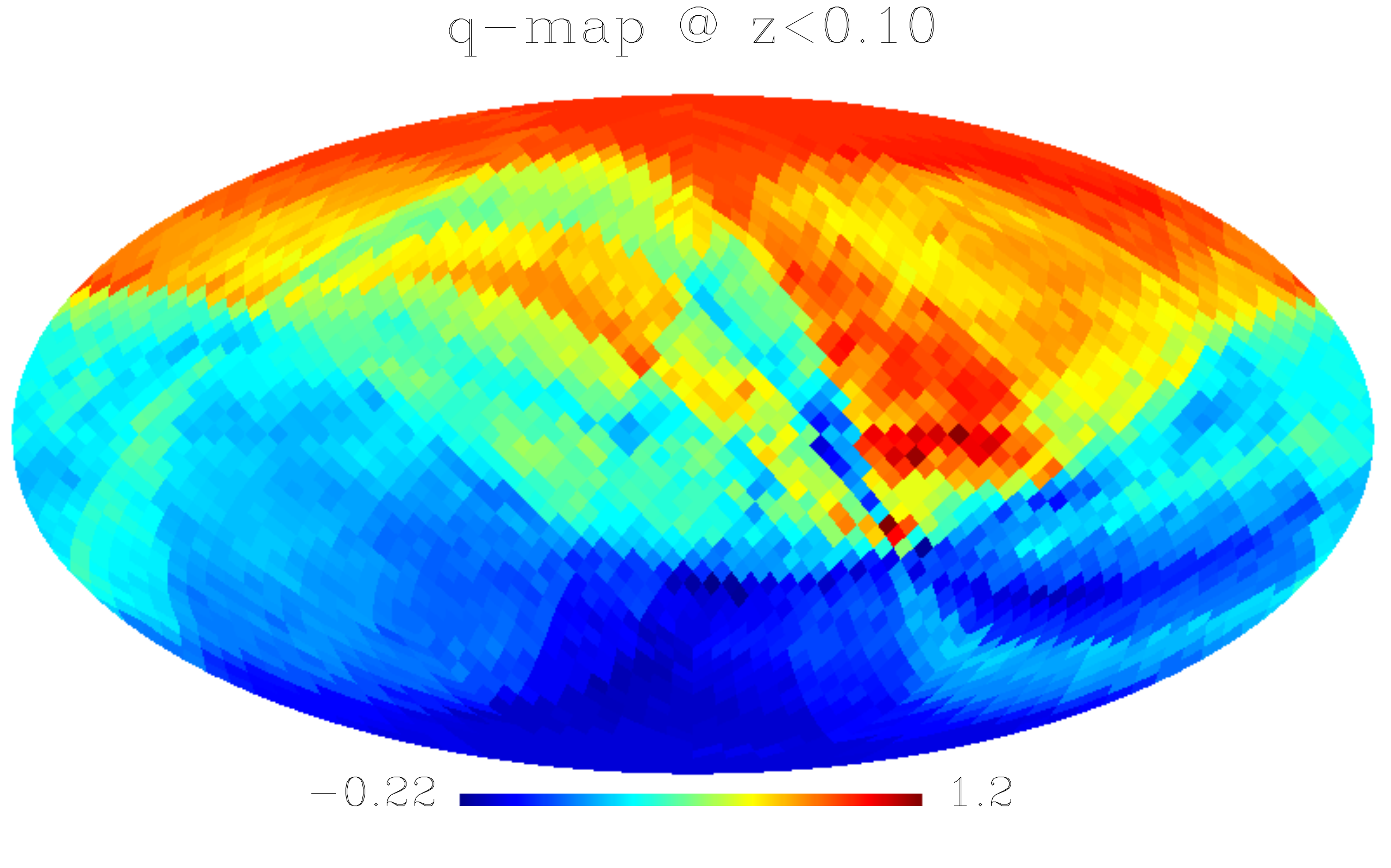}
\caption{The hubble-map of Pantheon SN at $z \leq 0.10$ (left panel), and the q-map at the same redshift range (right panel). We note that $h_0$ ranges from $0.685$ to $0.711$ in the left panel, while $q_0$ ranges from $-0.22$ to $1.23$ in the central one, }
\label{fig:maps_z0p1}
\end{figure*}

\begin{figure*}[t]
\includegraphics[width=0.30\textwidth, height=0.38\textheight, angle=90]{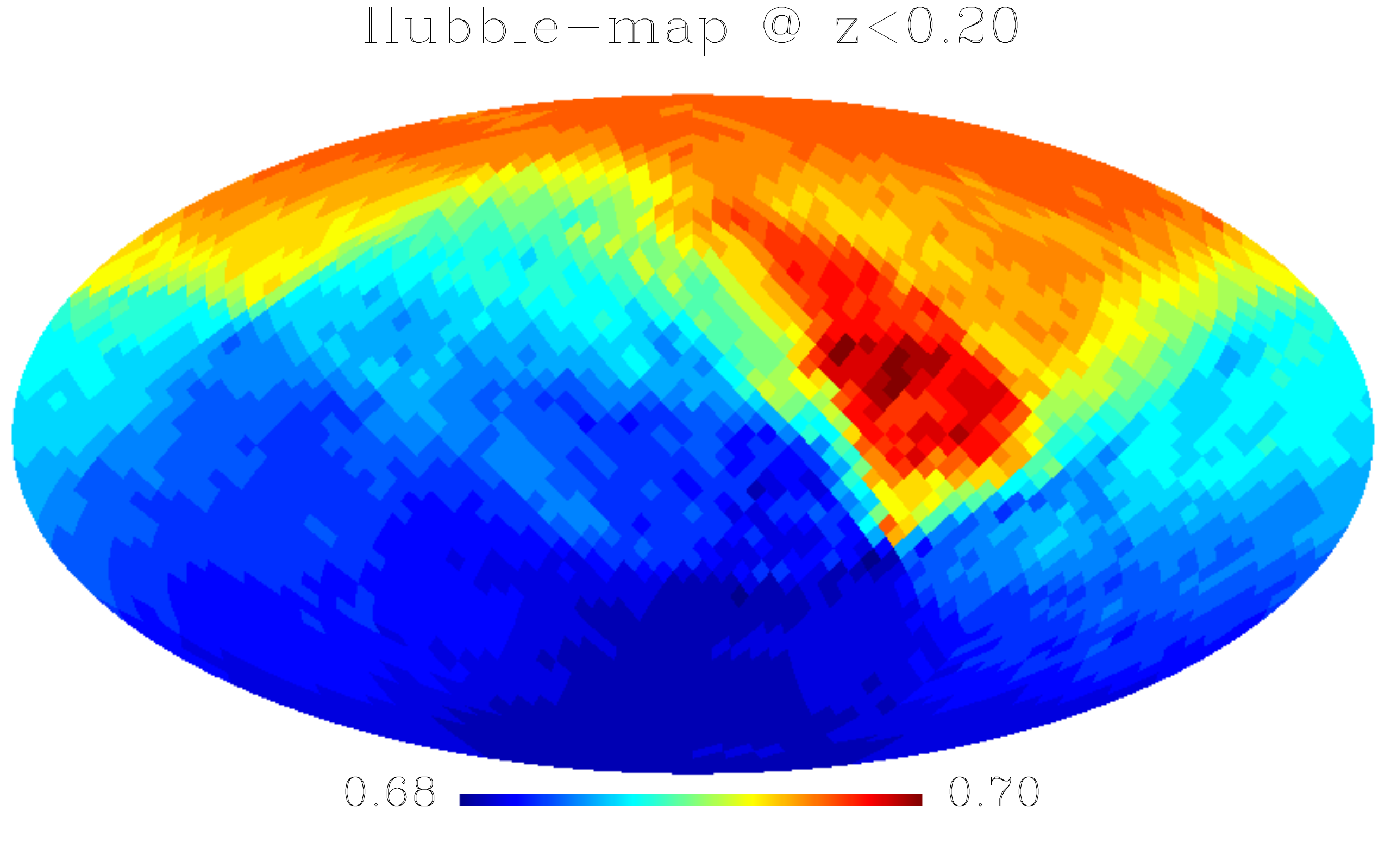}
\includegraphics[width=0.30\textwidth, height=0.38\textheight, angle=90]{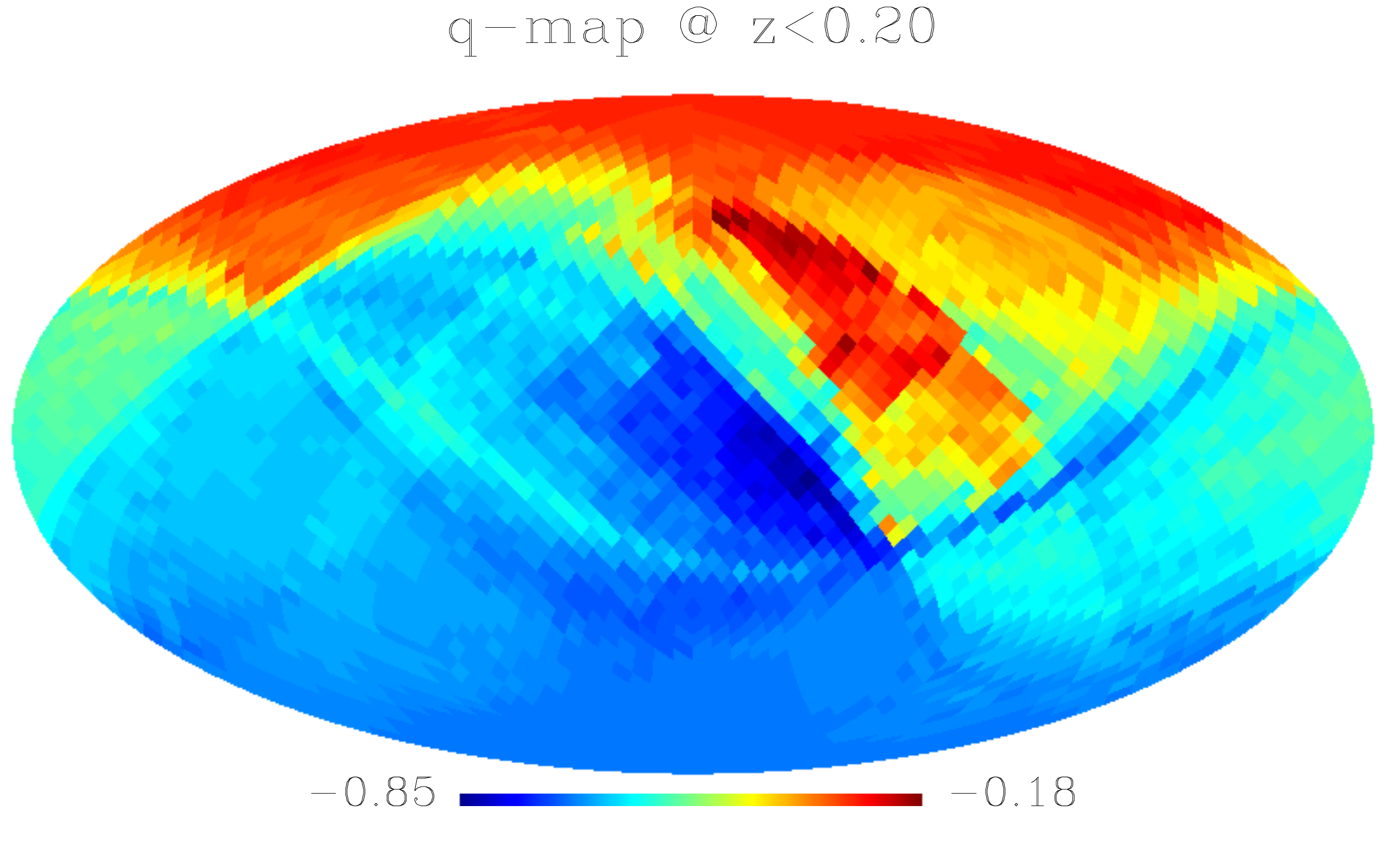}
\caption{The same as Fig.~(\ref{fig:maps_z0p1}) for the $z \leq 0.20$ SN Ia instead. We note that $h_0$ ranges from $0.676$ to $0.700$, $q_0$ ranges from $-0.85$ to $-0.18$.}
\label{fig:maps_z0p2}
\end{figure*}

\begin{figure*}[t]
\includegraphics[width=0.48\textwidth, height=0.23\textheight, angle=00]{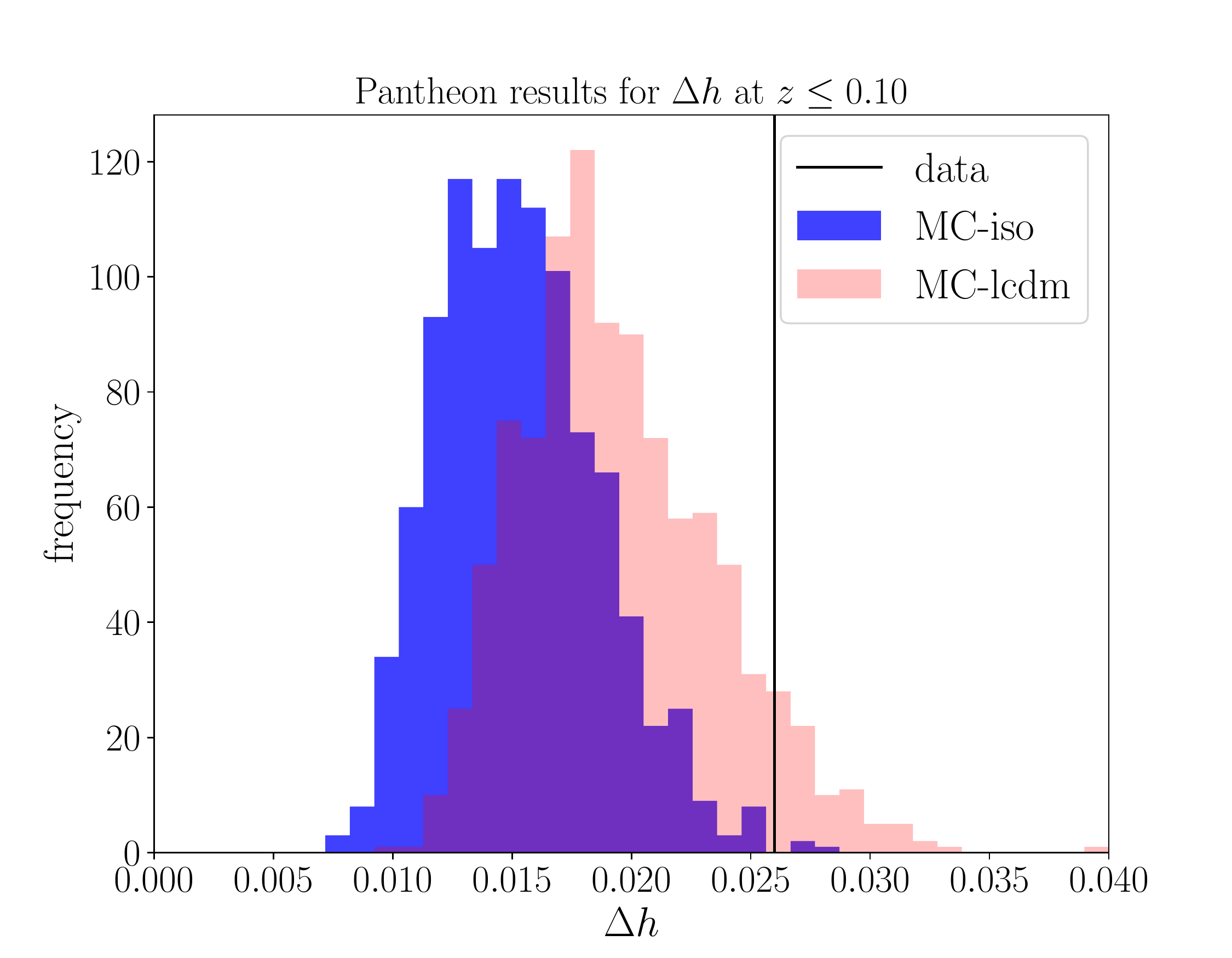}
\includegraphics[width=0.48\textwidth, height=0.23\textheight, angle=00]{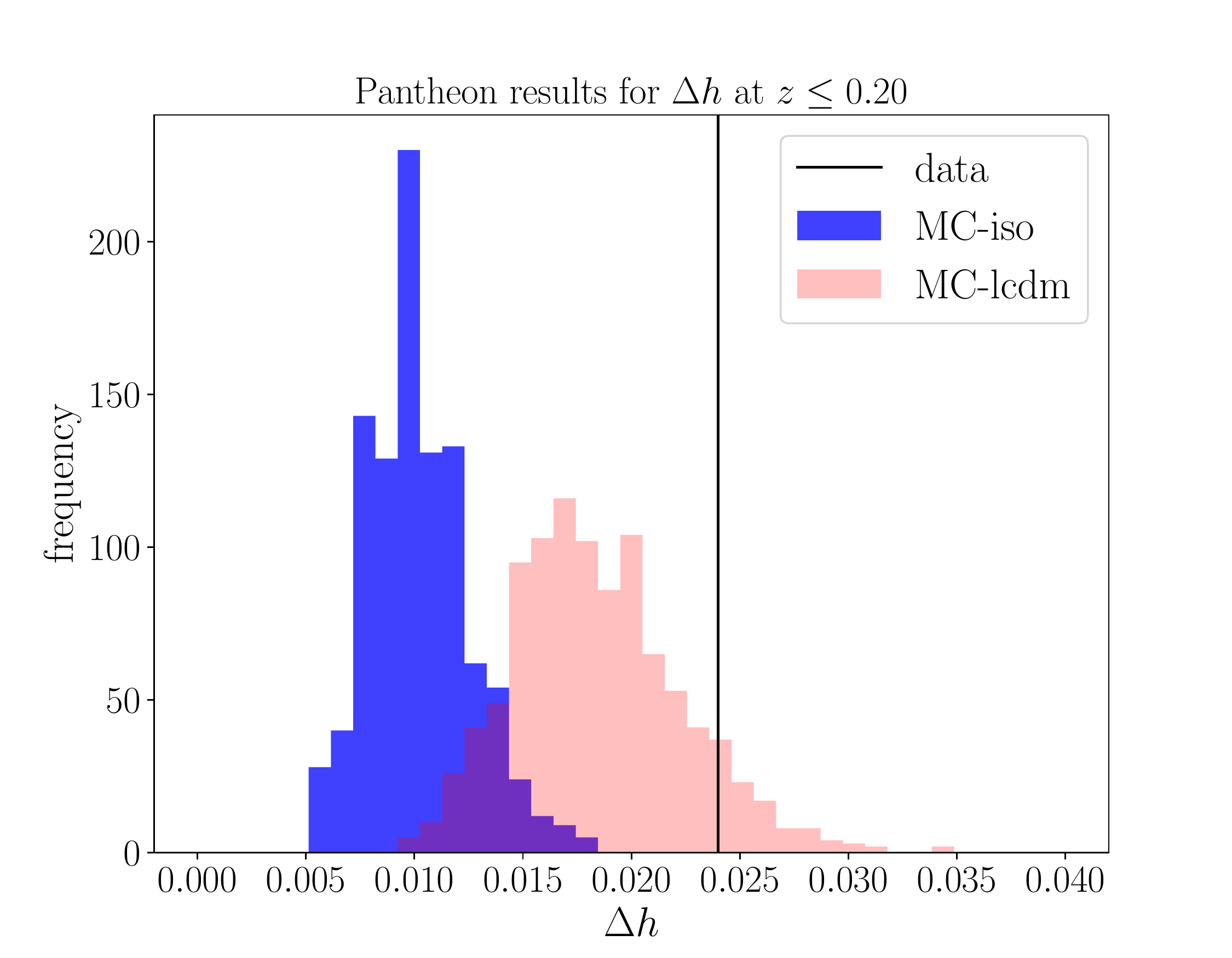}
\caption{The $\Delta h$ values of the {\it MC-lcdm} (pink histogram) and {\it MC-iso} (blue histogram) realizations compared to the actual data for $z \leq 0.1$ (left panel) and $z \leq 0.20$ (right panel), respectively. The vertical line corresponds to the real data result.}
\label{fig:MC_deltah}
\end{figure*}

\begin{figure*}[t]
\includegraphics[width=0.48\textwidth, height=0.23\textheight, angle=00]{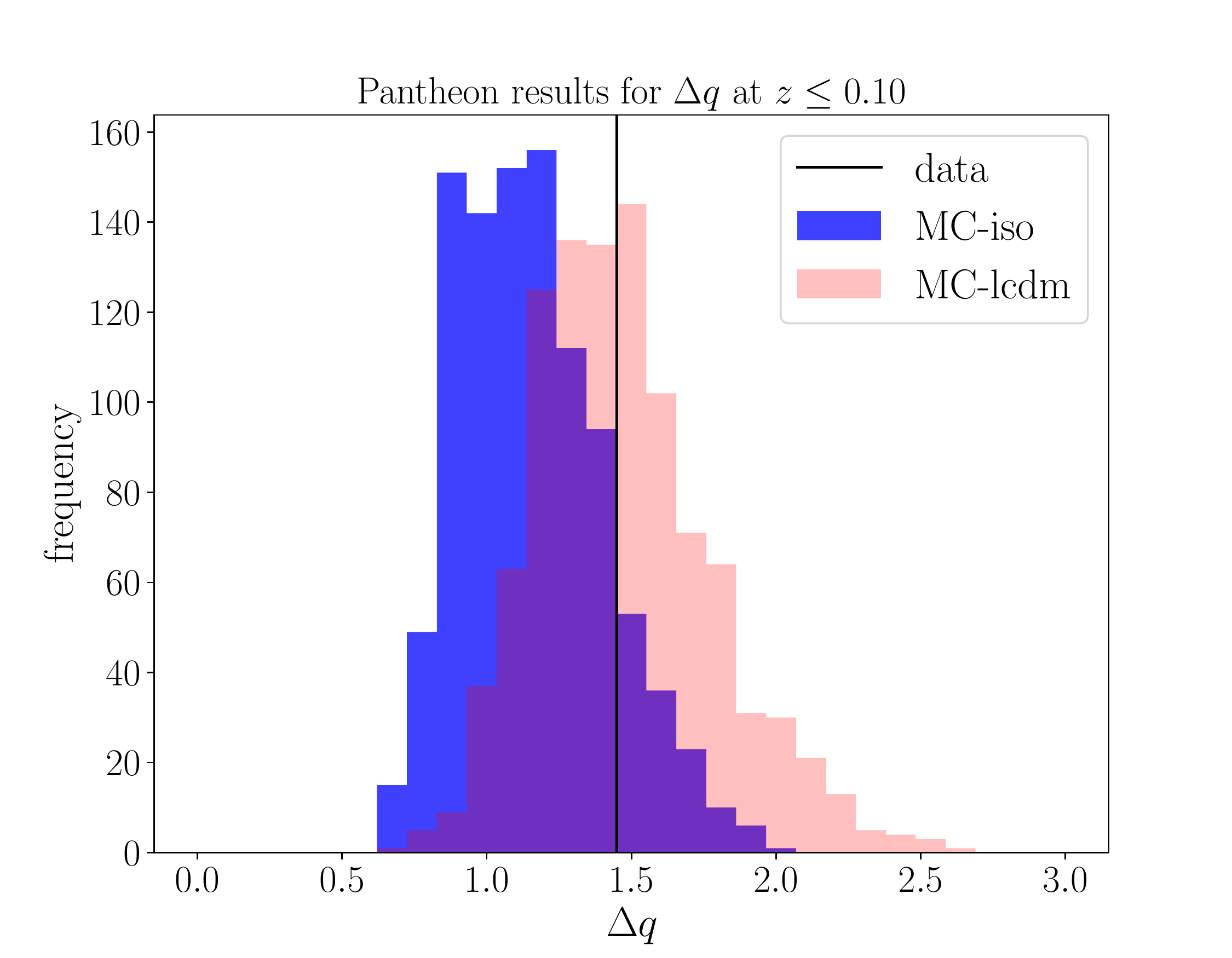}
\includegraphics[width=0.48\textwidth, height=0.23\textheight, angle=00]{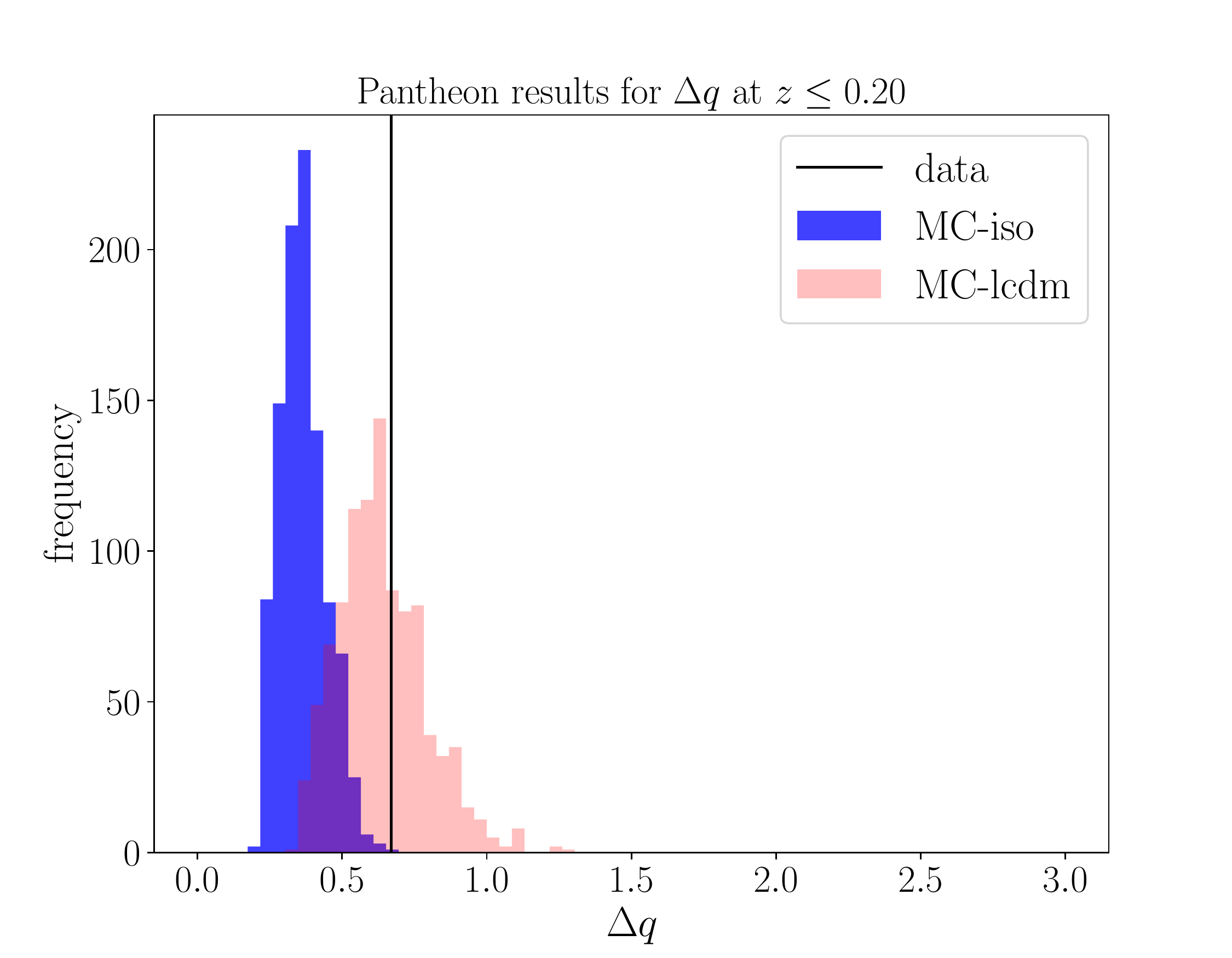}
\caption{The same as Fig.~\ref{fig:MC_deltah} for $\Delta q$.}
\label{fig:MC_deltaq}
\end{figure*}

\begin{table*}[t]
\begin{center}
\begin{tabular}{ccccc}
\hline
\hline
& & hubble-map \\ 
\hline
$z$ range & $\Delta h$ & {\it MC-iso} $p$-value (\%) & {\it MC-lcdm} $p$-value (\%) \\
\hline
$z \leq 0.10$ & $0.026$ & $0.3$ & $7.3$ \\
$z \leq 0.20$ & $0.024$ & $<0.1$ & $7.1$ \\
\hline
\hline
& & q-map \\
\hline
$z$ range & $\Delta q$ & {\it MC-iso} $p$-value (\%) & {\it MC-lcdm} $p$-value (\%) \\
\hline
$z \leq 0.10$ & $1.440$ & $37.4$ & $47.9$\\
$z \leq 0.20$ & $0.670$ & $0.1$ & $12.5$ \\
\hline
\hline 
\end{tabular}
\end{center}
\caption{Respectively: The redshift range, $\Delta h$, as defined in Eq.~(\ref{eq:delta}), and the $p$-values for each MC data-set. Below are shown the same quantities for the $q$-maps.}
\label{tab:tab_hubblemap_qmap} 
\end{table*}

We show the hubble-maps for both $z \leq 0.10$ and $z \leq 0.20$ in the left panels of Figures~\ref{fig:maps_z0p1} and~\ref{fig:maps_z0p2}, respectively. In the central panels of the same figures, we display the results for the $q$-maps in the same redshift intervals. As presented in Table~\ref{tab:tab_hubblemap_qmap}, the hubble-maps are in excellent agreement with previous results, such as $\Delta h = 0.030$ obtained by~\cite{2013A&A...553A..56K} using the Constitution compilation for $z \leq 0.20$, as well as $\Delta h = 0.023$ from Union2.1 for $z \leq 0.10$, as reported by~\cite{2016JCAP...04..036B}. If we allow for the $1\sigma$ error-bars in both $h^{\rm max}_{0}$ and $h^{\rm min}_0$ values, 
we then obtain $\Delta h = 0.026 \pm 0.086$ ($z \leq 0.10$) and $\Delta h = 0.024 \pm 0.058$ ($z \leq 0.20$), which is fully compatible with the $h_0$ uncertainty due to cosmic variance, that is, $\Delta_{h_0} = 0.015$ according to~\cite{2018arXiv180509900C} (see also~\citealt{2014PhRvL.112v1301B}). Regarding the $q$-maps, we find $\Delta q = 1.440$ at $z \leq 0.10$, and $\Delta q = 0.670$ at $z \leq 0.20$. Allowing again for the $1\sigma$ error-bars in $q^{\rm max}_{0}$ and $q^{\rm min}_0$, we obtain $\Delta q = 1.440 \pm 0.617$ and $\Delta q = 0.670 \pm 0.228$, respectively, at $z \leq 0.10$ and $z \leq 0.20$ redshift ranges\footnote{We emphasize that these $\Delta q$ are smaller than those reported in~\cite{2015ApJ...808...39B} because we did not marginalize over $h_0$. The same happens for $\Delta h$.}. We can clearly see that the highest $h_0$ and $q_0$ regions coincide with those hemispheres with the lowest SN Ia counts, as shown in Figures~\ref{fig:SN_maps_z0p1} and~\ref{fig:SN_maps_z0p2}. This potentially indicates a bias in both hubble- and $q$-maps due to it.   

We show the results of the MC analyses in Fig.~\ref{fig:MC_deltah} for the hubble-maps in both redshift intervals, whereas Fig.~\ref{fig:MC_deltaq} exhibits the results for the $q$-maps. For the hubble-maps, we can readily note that the {\it MC-lcdm} shows stronger agreement with the actual $\Delta h$ than the {\it MC-iso} realizations for both cases. This result shows that the $\Delta h$ obtained from the data is not unexpected in the $\Lambda$CDM scenario, and that its value is mostly due to the non-uniform celestial distribution of SN Ia, as suggested in the right panels of Figures~\ref{fig:maps_z0p1} and~\ref{fig:maps_z0p2}. This is also reflected in the $p$-values shown in the third and fourth columns of Table~\ref{tab:tab_hubblemap_qmap}, as very few isotropic MCs ($p$-value $<1$\%) gave $\Delta h \geq 0.026$, yet the $p$-value increases to $\sim 7$\% for the {\it MC-lcdm} case. Similar conclusions can be drawn for the q-maps, since the $p$-values significantly increase in the {\it MC-lcdm} realizations compared to the {\it MC-iso} ones, as depicted in both panels of Fig.~\ref{fig:MC_deltaq}, and in the last two columns of the lower part of Table~\ref{tab:tab_hubblemap_qmap}. In fact, the agreement between the real and simulated data-sets for the q-map is stronger than the hubble-map case at all redshift ranges. These results are also compatible with previous works, such as~\cite{2015ApJ...808...39B, 2016JCAP...04..036B}, in which the anisotropy found in the cosmological parameters were stronger than expected from simulations assuming a perfectly uniform distribution of objects. Hence, we conclude that there is null evidence for anomalous anisotropy in cosmic distances since most of the $\Delta h$ and $\Delta q$ are due to the high asymmetry of data points in the sky, and that they fully agree with simulations that assume the concordance model given the uncertainties of the real data.             

\section{Final Remarks} \label{sec:disc}

Following previous analyses~\citep{2015ApJ...808...39B, 2016JCAP...04..036B} we tested the cosmological isotropy using the recently released SN Ia compilation, i.e., the Pantheon data set. As in those analyses, we looked for signatures of isotropy departure by mapping the cosmographic parameters $h_0$ and $q_0$ across the sky, namely the hubble- and $q$-maps, and by estimating their statistical significance with MC simulations as well. For this purpose, we produced two sets of MCs: one that redistributes the SN in an uniform way, {\it MC-iso}, and another that assumes the cosmic distances to be given by the $\Lambda$CDM model within the error bars of the original data, {\it MC-lcdm}. We obtained $\Delta h = 0.026 \pm 0.086$ ($z \leq 0.10$) and $\Delta h = 0.024 \pm 0.058$ ($z \leq 0.20$), besides $\Delta q = 1.440 \pm 0.617$ and $\Delta q = 0.670 \pm 0.228$, also at $z \leq 0.10$ and $z \leq 0.20$, respectively. We found excellent agreement between the Pantheon results and previous analyses using older SN Ia compilations for $\Delta h$, whose value is also compatible with the cosmic variance estimative. Moreover, the {\it MC-lcdm} shows much stronger agreement with real data compared to the {\it MC-iso} for both hubble- and $q$-maps. 

Finally, it is important to emphasize that the anisotropies herein reported are mostly due to non-uniform angular distribution of SN Ia, thus a selection effect, rather than a real departure of the concordance model. Therefore, the FLRW Universe can describe low-$z$ SN Ia observations very well, which guarantees the validity of the CP. In future works, different methods to test the cosmological isotropy assumption, as those presented in~\citep{marinucci2011, jun2012, hitczenko2012, guinness2016, sahoo2017}, can be readily envisaged and performed in light of larger, more homogeneous cosmological data sets, that will be provided in the years to come. Among potential applications, we list the SN Ia sample from LSST~\citep{2009arXiv0912.0201L}, luminosity distance measurements from standard sirens with next-generation gravitational wave experiments~\citep{2018PhRvD..97j3005C, 2018EPJC...78..356L}, besides large-area galaxy surveys such as SKA~\citep{2015aska.confE..32S}. Given all these observational data and statistical machinery available, we should be able to definitely underpin the validity of CP at large scales.  

\acknowledgments
We thank Dan Scolnic for kindly providing us the Pantheon compilation. UA acknowledges financial support from CAPES. CAPB acknowledges financial support from the South African SKA Project. JSA acknowledges support from CNPq (Grants no. 310790/2014-0 and 400471/2014-0) and FAPERJ (Grant no. 204282). BS is supported by the DTI-PCI program of the Brazilian Ministry of Science and Technology. Some of the analyses herein performed used the {\sc HEALPix} software package.

\end{document}